\documentstyle[12pt,epsfig]{article}

\newcommand{\f}{\frac}
\newcommand{\non}{\nonumber \\}
\newcommand {\beq}{\begin{equation}}
\newcommand {\eeq}{\end{equation}}
\newcommand {\beqa}{\begin{eqnarray}}
\newcommand {\beqal}{\begin{eqnarray}\label}
\newcommand {\eeqa}{\end{eqnarray}}
\newcommand {\bc}{\begin{center}}
\newcommand {\ec}{\end{center}}
\newcommand {\s}{\sigma}
\newcommand {\al}{\alpha}

\newcommand{\ket}[1]{\left | #1 \right\rangle}
\newcommand{\expect}[1]{\left\langle #1 \right\rangle}

\def\nn{\nonumber}

\def\vs5{\vspace*{5mm}}
\def\vs1{\vspace*{1cm}}
\def\vs2{\vspace*{2cm}}
\def\hs5{\vspace*{5mm}}
\def\hs1{\hspace*{1cm}}

\begin{document}

\title{
\hfill\parbox{4cm}{\normalsize IMSC/2003/09/28}\\
\vspace{2cm}
Closed String Tachyons on $C/Z_N$
\author{S. Sarkar and B. Sathiapalan\\ 
\\
{\em Institute of 
Mathematical Sciences}\\
{\em Taramani}\\{\em Chennai, India 600113}}}
\maketitle

\begin{abstract}

We analyse the condensation of closed string tachyons on the $C/Z_N$ 
orbifold. We construct the potential for the tachyons upto the quartic 
interaction term in the large $N$ limit. In this limit there are near 
marginal tachyons. The quartic coupling for these tachyons is calculated 
by subtracting from the string theory amplitude for the tachyons, the 
contributions from the massless exchanges, computed from the effective  
field theory. We argue that higher point interaction terms are 
are also of the same order in $1/N$ as the quartic term and are necessary 
for existence of the minimum of the tachyon potential that is consistent 
with earlier analysis.

\end{abstract}

\newpage

\section{Introduction}

The condensation of closed string tachyons has been studied in some 
specific models. For closed string on nonsupersymmetric $C/Z_N$ orbifold, 
there are tachyons 
in the twisted sectors. Being in the twisted sectors, 
these tachyons are localised at the fixed point of the orbifold. This 
gives a more manageable set up for the study of condensation of 
these tachyons. In \cite{pol} D-brane probes were used to follow this 
process 
and it was shown that the end result of the condensation with generic 
tachyonic perturbations is flat space. There are however specific 
perturbations which drive $C/Z_N$ to $C/Z_{N-2}$ or other lower 
nonsupersymmetric orbifolds \cite{vafa1,harvey1}. Condensation of tachyons 
for a more general 
background namely the twisted circles confirms these observations 
\cite{justin1}.   
For other studies on condensation of localized closed string tachyons 
see \cite{atish1,atish3,other,dine}.   

In this paper we consider the problem of tachyon 
condensation for Type II theory on the $C/Z_N$ orbifold in the large $N$ 
limit. In this limit there are tachyons which become almost marginal  
and it makes sense to write an effective action involving the tachyons and 
the other massless particles (graviton, dilaton) while integrating out the 
massive string modes.
The aim is to compute the effective tachyon potential for large $N$. 
With this as the guideline, we construct the tachyon potential upto the 
quartic interaction term. The procedure followed is along the lines of 
\cite{narain}. The four point amplitude for the twisted sector tachyons 
(of $m^2=-1/N$) is first 
calculated \cite{cftorb}. In the large $N$ limit when the tachyons are 
nearly 
massless we take the zero momentum limit of this amplitude. The 
nonderivative quartic coupling for the tachyons is then obtained by 
subtracting the contribution of the massless exchanges from the string 
four point amplitude. The four point amplitude with massless exchanges, 
which in this case are the graviton and the dilaton is obtained from the 
low energy effective field theory of tachyons coupled to these massless 
fields.

The quartic coupling for the tachyon potential is found to be of the order 
$1/N$. We get the the height of the potential 
to the lowest order in $1/N$. 
We expect this minimum to correspond to the $C/Z_{N-k}$ orbifold. 

However there are various points which show that the higher point 
interaction couplings are also comparable to the quartic coupling. One 
being that, with a quartic potential having global $O(2)$ symmetry, 
we expect a 
particle of positive $(mass)^2=-2m^2$, where $m^2$ is the mass of the 
tachyon, and the 
usual Goldstone boson. However these modes are not present in the 
spectrum of closed string on $C/Z_{N-k}$. 
Furthermore if we stick to the predicted 
height of the tachyon potential \cite{atish2}, we find that there is a 
mismatch by a factor of $1/N$ in the height of the potential, when the 
minimum is expected to correspond to the $C/Z_{N-k}$ orbifold. 
However since the above modes are absent in the 
tree-level spectrum of closed string on $C/Z_{N-k}$ orbifold we conclude 
that the higher point amplitudes are also of the order $1/N$.
This includes the term $\phi^N$ allowed by the twist symmetry. These 
higher order terms modify the potential and the spectrum already to the 
lowest order in $1/N$.

Furthermore, the four point coupling is subject to field redefinitions. 
One can make the offshell contact term as large or small as one wishes and 
can also change the sign. This makes the coupling nonuniversal and the 
existence of the minimum is not clear in this approach if one truncates 
the potential upto the quartic term.
We elaborate on these points in section 5.

This paper is organised as follows. In section 2, we compute the spectrum 
for closed strings on $C/Z_N$ orbifold and show that there are tachyons. 
In section 3, using the conformal field theory of $C/Z_N$ orbifold we 
review the calculation of the four point amplitude of the tachyons in the 
large $N$ limit where the tachyon is nearly marginal. In 
section 4, we find the OPE of two tachyon vertices and show that the only 
intermediate massless exchanges are the graviton and dilaton. In section 
5, from the effective field theory of the tachyon coupled to the graviton 
and dilaton we compute the exact contribution of these massless exchanges 
to the four point tachyon amplitude. The quartic coupling for the tachyon 
is then obtained after subtraction of the massless exchanges from the 
string theory amplitude and write down the potential upto the quartic 
interaction term. We give our conclusions in section 6.  

\section{Closed string spectrum}

We compute the spectrum of the closed string on the $C/Z_N$ 
orbifold. We will be concerned with the NS sector as it is in this sector 
that tachyons appear. In the RNS formulation of superstring, we consider 
$x^{\mu}$ and $\Psi^{\mu}$ as world sheet fields corresponding to the 
nonorbifolded directions. For the orbifolded directions we have the 
complex $X$,$\bar{X}$,$\psi$,$\bar{\psi}$ fields as defined below,

\beqa
X=X^8+iX^9 \mbox{,\hspace{0.5cm}} \bar{X}=X^8-iX^9
\mbox{;\hspace{0.5cm}}\psi=\psi^8+i\psi^9 \mbox{,\hspace{0.5cm}} 
\bar{\psi}=\psi^8-i\psi^9
\eeqa

The ${Z}_N$ group action on ${C}$ defines the following 
boundary conditions on $X$,$\bar{X}$ and $\psi$, 
$\bar{\psi}$ for the closed string in the NS sector,

\beqa
X(\s+2\pi,\tau)&=&e^{2\pi i \f{k}{N}}X(\s,\tau)\non
\bar{X}(\s+2\pi,\tau)&=&e^{-2\pi i \f{k}{N}}\bar{X}(\s,\tau)\non
\psi(\s+2\pi,\tau)&=&e^{2\pi i (\f{k}{N}+\f{1}{2})}\psi(\s,\tau)\non
\bar{\psi}(\s+2\pi,\tau)&=&e^{-2\pi i 
(\f{k}{N}-\f{1}{2})}\bar{\psi}(\s,\tau)
\eeqa

These boundary conditions give the following mode expansions for the 
world sheet scalars,
\footnote{$\alpha^{'}=2$ in all  
calculations},

\beqa
\partial_{z}X(z)=-i\sum_{m=-\infty}^{\infty} 
 \f{\al_{m-\f{k}{N}}}{z^{m+1-\f{k}{N}}}
\mbox{\hspace{0.5cm}}
\partial_{z}\bar{X}(z)=-i\sum_{m=-\infty}^{\infty}
 \f{\bar{\al}_{m+\f{k}{N}}}{z^{m+1+\f{k}{N}}}
\eeqa

\beqa
\partial_{\bar{z}}X(\bar{z})=-i\sum_{m=-\infty}^{\infty}
 \f{\tilde{\al}_{m+\f{k}{N}}}{\bar{z}^{m+1+\f{k}{N}}}
\mbox{\hspace{0.5cm}}
\partial_{\bar{z}}\bar{X}(\bar{z})=-i\sum_{m=-\infty}^{\infty}
 \f{\tilde{\bar{\al}}_{m-\f{k}{N}}}{\bar{z}^{m+1-\f{k}{N}}}
\eeqa

and for the fermions,

\beqa
\psi(z)=\sum_{r\in Z+\f{1}{2}}
 \f{\psi_{r-\f{k}{N}}}{z^{r+\f{1}{2}-\f{k}{N}}}
\mbox{\hspace{0.5cm}}
\bar{\psi}(z)=\sum_{r \in Z+\f{1}{2}}
 \f{\bar{\psi}_{r+\f{k}{N}}}{z^{r+\f{1}{2}+\f{k}{N}}}
\eeqa

\beqa
\psi(\bar{z})=\sum_{r\in Z+\f{1}{2}}
\f{\tilde{\psi}_{r+\f{k}{N}}}{\bar{z}^{r+\f{1}{2}+\f{k}{N}}}
\mbox{\hspace{0.5cm}}
\bar{\psi}(\bar{z})=\sum_{r \in Z+\f{1}{2}}
 \f{\tilde{\bar{\psi}}_{r-\f{k}{N}}}{\bar{z}^{r+\f{1}{2}-\f{k}{N}}}
\eeqa

From the usual OPEs of the world sheet fields and the mode expansions we 
get the following cannonical commutation relations,

\beqa
[\al_{m-\f{k}{N}},\bar{\al}_{n+\f{k}{N}}]=(m-\f{k}{N})\delta_{m+n,0}
\mbox{\hspace{0.5cm}} 
[\al_m^{\mu},\al_n^{\nu}]=m\eta^{\mu\nu}\delta_{m+n,0}
\eeqa

\beqa
[\tilde{\al}_{m+\f{k}{N}},\tilde{\bar{\al}}_{n-\f{k}{N}}]=
(m+\f{k}{N})\delta_{m+n,0}
\mbox{\hspace{0.5cm}}
[\tilde{\al}_m^{\mu},\tilde{\al}_n^{\nu}]=m\eta^{\mu\nu}\delta_{m+n,0}
\eeqa

\beqa
\{\psi_{r-\f{k}{N}},\bar{\psi}_{s+\f{k}{N}}\}=\delta_{r+s,0}
\mbox{\hspace{0.5cm}}
\{\psi_r^{\mu},\psi_s^{\nu}\}=\eta^{\mu\nu}\delta_{r+s,0}
\eeqa

\beqa
\{\tilde{\psi}_{r+\f{k}{N}},\tilde{\bar{\psi}}_{s-\f{k}{N}}\}=\delta_{r+s,0}
\mbox{\hspace{0.5cm}}
\{\tilde{\psi}_r^{\mu},\tilde{\psi}_s^{\nu}\}=\eta^{\mu\nu}\delta_{r+s,0}
\eeqa

The holomorphic part of the energy momentum tensor for the world sheet 
fields corresponding to the orbifolded directions is given by,

\beqa
T_{zz}(z)=-\partial X \partial 
\bar{X}-\f{1}{2}\bar{\psi}\partial{\psi}
-\f{1}{2}\psi\partial{\bar{\psi}}
\eeqa 

Using the mode expansions and the cannonical commutation relations one 
gets the zero mode part of the energy momentum tensor,

\beqal{l0}
L_0&=&\sum_{n=0}^{\infty}[\al_{-n-\f{k}{N}}\bar{\al}_{n+\f{k}{N}}+\f{1}{2} 
(n+\f{k}{N})] 
+ \sum_{n=1}^{\infty}[\bar{\al}_{-n+\f{k}{N}}\al_{n-\f{k}{N}}+\f{1}{2}
(n-\f{k}{N})]\non
&+&\sum_{r\geq 1/2}
[(r+\f{k}{N})\psi_{-r-\f{k}{N}}\bar{\psi}_{r+\f{k}{N}}+
(r-\f{k}{N})\bar{\psi}_{-r+\f{k}{N}}\psi_{r-\f{k}{N}}-r]
\eeqa

Where we have included the contributions from the normal ordering 
constant. The contribution to the zero point energy from the fields on 
the orbifolded complex plane is now given by,

\beqa
E_{orb}&=&\f{1}{2}\f{k}{N}+\sum_{n=0}^{\infty} n-\sum_{r\geq 1/2} r \non
&=& \f{1}{2}\f{k}{N}-\f{1}{8}
\eeqa

In NS-sector in the light cone gauge, we have six real periodic bosons and 
six antiperiodic fermions which contributes an amount of $-3/8$ to the 
zero point energy.
Adding the zero point energies for all the directions, for the left moving 
part, we get,

\beqa
E_L=-\f{1}{2}(1-\f{k}{N})
\eeqa

It may seem that for  $ 1/2 < k/N < 1$, the $r=1/2$ term from the second 
fermionic 
part in (\ref{l0}) contributes an additional ($1/2-k/N$) after normal 
ordering. Therefore, for this case we should have,

\beqa
E_L =-\f{k}{2N}  
\eeqa

However, since normal ordering is only a prescription for removing 
infinities from the zero point energy, we must choose one so that the zero 
point energy is consistent with that which is obtained from the 
world sheet conformal field theory. It will be seen in the next 
section, that a choice of the prescription, where we keep the $r=1/2$ 
as it is, is consistent with the dimensions of the twist operators. The 
mass spectrum is thus given by,

\beqa
M^2&=&n+\tilde{n}-(1-\f{k}{N})  \mbox{\hspace{.2cm} for 
\hspace{.2cm}}
0<k/N<1\non
\eeqa

Where $n$  and $\tilde{n}$ are level numbers which are no longer integers
for the twisted sectors.
The ground state of the NS sector is thus tachyonic. 
For the $(N-k)th$ sector we have a tachyon of $(mass)^2= -k/N$.
Some of the excited states are also tachyonic their masses are given by,

\beqal{marginal}
\bar{\psi}_{-\f{1}{2}+\f{k}{N}}\ket{0}&\equiv & -\f{k}{N}
\mbox{\hspace{.1cm}:\hspace{.1cm}
marginal for,\hspace{.2cm}} N \rightarrow \infty \\
\al_{-\f{k}{N}}\ket{0}&\equiv & -(1-\f{3k}{N}) 
\mbox{\hspace{.1cm}:\hspace{.1cm}
tachyonic for,\hspace{.2cm}} 3k<N
\eeqa

From the moding of the oscillators it can be 
seen that there are no massless states in the twisted sector, as the 
oscillators are moded by $k/N$ and the zero point energy is $-k/2N$. The 
massless states arise from the untwisted sector and these are the usual 
graviton, dilaton and the antisymmetric second rank tensor. In the 
following sections we will construct the tachyon potential in the large 
$N$ limit where the tachyons become marginal.

\section{Four point amplitude from CFT} 

In this section we review the computation of the four point
amplitude for the tachyons that we found in the 
spectrum \cite{cftorb}. This gives the four point tachyon amplitude with 
all the 
massless and the massive exchanges. In the next section we will compute 
the exact contribution from the massless exchanges. Subtracting this from 
the amplitude computed here gives the effective four point coupling for 
the tachyon field.

The vacuum for the twisted sector that is labeled by $k/N$ is created 
from the untwisted vacuum by the action of the bosonic twist fields, 
$\s_{\pm\f{k}{N}}$ and the fermionic twist fields $s_{\pm\f{k}{N}}$.

The OPEs of these twist fields with the world sheet fields, 
$X,\bar{X},\psi,\bar{\psi}$ are given by \cite{cftorb},

\beqal{ope}
\partial_z X(z) \s_+(w,\bar{w}) &\sim& (z-w)^{-(1-\f{k}{N})}\tau_+(w,{\bar
w})\non
\partial_z {\bar X}(z) \s_+(w,\bar{w}) &\sim&
(z-w)^{-\f{k}{N}}\tau^{'}_{+}(w,{\bar w})\non
\partial_{\bar z} X({\bar z}) \s_+(w,\bar{w}) &\sim&
({\bar z}-{\bar w})^{-\f{k}{N}}{\tilde \tau}^{'}_{+}(w,{\bar
w})\non
\partial_{\bar z} {\bar X}({\bar z}) \s_+(w,\bar{w}) &\sim&
({\bar z}-{\bar w})^{-(1-\f{k}{N})}{\tilde \tau}_{+}(w,{\bar
w})
\eeqa

where, $\tau_+$, $\tau_+^{'}$, $\tilde{\tau}_+$, $\tilde{\tau}_+^{'}$ are 
excited twist fields. Using (\ref{ope}) 
and 
the OPE of the twist fields with the energy momentum tensor, the world 
sheet dimension of the bosonic twist fields are found to be,

\beqa
h_{\s}=\bar{h}_{\s}=\f{1}{2}\f{k}{N}(1-\f{k}{N})
\eeqa

Bosonising the world sheet fermions,

\beqa
\psi(z) &=& -i\sqrt{2} e^{iH(z)}\\
{\bar \psi}(z) &=& -i\sqrt{2} e^{-i H(z)}
\eeqa

giving the fermionic twist fields as,
$s_{\pm}=e^{\pm i\f{k}{N}H(z)}$. From this we get the  following OPEs of 
the fermionic fields with the 
twist fields.

\beqa
{\bar \psi}(z)s_{+}(w) &=&  -i\sqrt{2} e^{-iH(z)}e^{i\f{k}{N}H(w)}\non
&\sim & -i\sqrt{2}
(z-w)^{-\f{k}{N}}e^{-i(1-\f{k}{N})H(z)}[1+(z-w)\partial H(z)]
\eeqa

Similarly,

\beqa
\psi(z)s_{+}(w) &=&  -i\sqrt{2} e^{iH(z)}e^{i\f{k}{N}H(w)}\non
&\sim & -i\sqrt{2}
(z-w)^{\f{k}{N}}e^{i(1+\f{k}{N})H(z)}[1+(z-w)\partial H(z)]
\eeqa

For the  fermionic string, vertices  for the twist fields in the $(-1,-1)$ 
and $(0,0)$ pictures are given by,

\beqa
V_{(-1,-1)}^{+} (z,{\bar z})
&=& e^{-\phi}e^{-\tilde{\phi}}{\tilde s}_+ s_+ \s_+ 
e^{ik.x}(z,{\bar z})\\
V_{(0,0)}^{+} (z,{\bar z})&=& e^{\phi} T_F e^{\tilde{\phi}} {\tilde{T}}_F 
V_{(-1,-1)}^{+} (z,{\bar z})
\eeqa

Where,

\beqa
T_F=-\f{1}{4}(\partial X {\bar \psi} + \partial {\bar X} 
\psi)-\f{1}{2}\partial x.\Psi
\eeqa

Note that the dimension of the vertex gives the mass of the 
tachyon,

\beqa
M^2=-(1-\f{k}{N})
\eeqa

This corresponds to the ground state tachyons in the twisted sector.
For the near marginal tachyons, in the large $N$ limit, which are in 
the $(N-k)th$ sector, the vertex operator in the $(-1,-1)$ picture is,

\beqa
V_{(-1,-1)}^{+} (z,{\bar z})
= 
e^{-\phi}e^{-\tilde{\phi}}e^{i(1-\f{k}{N})H(z)} 
e^{-i(1-\f{k}{N})\tilde{H}(\bar{z})}
\s_+e^{ik.x}(z,{\bar z}) 
\eeqa

The four point amplitude for these lowest lying tachyons can now be 
computed by taking two vertices in the $(0,0)$ picture and two in the 
$(-1,-1)$ picture. 

\beqal{expect}
C\int_C d^2z\expect{V_{(-1,-1)}^{-}(z_{\infty},{\bar z}_{\infty})
e^{\phi} T_F e^{\tilde{\phi}} {\tilde{T}}_F V_{(-1,-1)}^{+} (1) 
V_{(-1,-1)}^{-}(z,{\bar z})
e^{\phi} T_F e^{\tilde{\phi}} {\tilde{T}}_F
V_{(-1,-1)}^{+} (0)}
\eeqa

\noindent
The constant $C=g_{c}^4C_{s}^2$. Where $C_{s}^2$ is related to $g_c$ by,

\beqa
C_{s}^2=\f{4\pi}{g_{c}^2}
\eeqa

\noindent
This amplitude can now be  computed and is given by \cite{cftorb},

\beqa
I= C(k_1.k_3)^2 \int_{C} d^2 z  \f{|z|^{-2-s} |1-z|^{-2-t}}{|F(z)|^2}
\eeqa

\noindent
Where, $F(z)$ is the hypergeometric function,

\beqa
F(z)\equiv F(\f{k}{N},1-\f{k}{N};1;z)=\f{1}{\pi}\int_{0}^{1}dy 
y^{-\f{k}{N}}(1-y)^{-(1-\f{k}{N})}(1-yz)^{-\f{k}{N}}
\eeqa

and, 
$s=-(k_1+k_2)^2$,  $t=-(k_2+k_3)^2$,  $s=-(k_3+k_1)^2$. 

In the large $N$ approximation, 

\beqal{fz}
F(z) \sim 1+ \f{k}{N}(z+\f{1}{2}z^2+\f{1}{3}z^3+...)+O((k/N)^2)  
\eeqa

Note that the terms proportional to $k/N$ in (\ref{fz})
shift the s-channel pole. There 
is an additional factor of $(k_1.k_2)^2$, due to which the contact term 
from any of the terms of (\ref{fz}) apart from $1$, would atleast be of 
$O((k/N)^2)$. With this observation, the integral can now be performed for 
$F(z) \rightarrow 1$.

\beqal{4pt}
I&=&C2\pi (k_1.k_3)^2\f {\Gamma (-\f{s}{2})\Gamma (-\f{t}{2}) \Gamma 
(1+\f{s}{2}+\f{t}{2})}{\Gamma (-\f{s}{2}-\f{t}{2}) \Gamma (1+\f{s}{2}) 
\Gamma (1+\f{t}{2})}\non
&=&-(4\pi)^2 g_{c}^2 \times \f{1}{4} (u-2m^2)^2(\f{1}{s}+\f{1}{t})
\f {\Gamma (1-\f{s}{2})\Gamma (1-\f{t}{2}) \Gamma
(1+\f{s}{2}+\f{t}{2})}{\Gamma (1-\f{s}{2}-\f{t}{2}) \Gamma (1+\f{s}{2})   
\Gamma (1+\f{t}{2})}
\eeqa

Now using $s+t+u=4m^2$,

\beqal{4pt1}
I=-4\pi^2 g_{c}^2
[\f{(t-2m^2)^2}{s}+\f{(s-2m^2)^2}{t}&+&3(s+t)-8m^2]\\
&\times& \f {\Gamma (1-\f{s}{2})\Gamma (1-\f{t}{2}) \Gamma
(1+\f{s}{2}+\f{t}{2})}{\Gamma (1-\f{s}{2}-\f{t}{2}) \Gamma (1+\f{s}{2})
\Gamma (1+\f{t}{2})}\nn 
\eeqa

We have to expand the gamma funcions. Now since we are interested in the 
order $O(\f{1}{N})$ 
of the amplitude, any correction to the exapnsion of the gamma functions 
to that when the limit  $s, t \rightarrow 0$ is taken, will be atleast of 
order $O(\f{1}{N})$. But the factor multiplying the gamma functions is 
already of order $O(\f{1}{N})$ except for the pole terms. So we can take 
the contribution of the gamma functions to be $1$.  
Thus we can write the string amplitude in the 
zero momentum limit as,

\beqal{string4pt}
I &\sim& -4\pi^2 g_{c}^2
[\f{(t-2m^2)^2}{s}+\f{(s-2m^2)^2}{t}+3(s+t)-8m^2] 
\eeqa

\section{OPE of two tachyon vertices}

In this section we compute the OPE of two tachyon vertices with one in the 
$(0,0)$ picture and another in the  $(-1,-1)$ picture and find the 
couplings of tachyon to the massless particles. 
\\

The OPE we wish to find is,

\beqal{ope2}
V_{(0,0)}^{-} (z,{\bar z})V_{(-1,-1)}^{+} (w,{\bar w})&=&
[\f{1}{4}(\partial X {\bar \psi} + \partial {\bar X}
\psi)+\f{1}{2}\partial x.\Psi][\f{1}{4}({\bar \partial} X {\tilde {\bar 
\psi}} + 
{\bar \partial} {\bar X}
{\tilde \psi})+\f{1}{2}{\bar{\partial} x}.{\tilde \Psi}]\non
\times {\tilde s}_- s_-\s_- e^{ik.x}(z,{\bar z})
& \times & 
e^{-\phi}e^{-\tilde{\phi}}{\tilde s}_+ s_+ \s_+
e^{ip.x}(w,{\bar w})\non
&=&[\f{1}{4}(\partial X {\bar \psi} + \partial {\bar X}
\psi)+k.\Psi][\f{1}{4}({\bar \partial} X {\tilde {\bar \psi}} +
{\bar \partial} {\bar X}
{\tilde \psi})+ k.{\tilde \Psi}]\non
\times {\tilde s}_- s_- \s_-e^{ik.x}(z,{\bar z})
& \times &
e^{-\phi}e^{-\tilde{\phi}}{\tilde s}_+ s_+ \s_+
e^{ip.x}(w,{\bar w})
\eeqa

\noindent
Now, the following OPEs are necessary to compute (\ref{ope2}).

\beqa
e^{ik.x}(z,{\bar z})e^{ip.x}(w,{\bar w}) &\sim & 
|z-w|^{2k.p}e^{i(p+k).x}(w,{\bar w})[1+(z-w)(k-p)_{\mu}\partial x^{\mu} 
\\
&+& ({\bar z}-{\bar w})(k-p)_{\mu}{\bar \partial} x^{\mu}  + 
|z-w|^{2}(k-p)_{\mu} 
(k-p)_{\nu} \partial 
x^{\mu} {\bar \partial} x^{\mu}]\nonumber
\eeqa 

\noindent
For the fermionic twist fields,

\beqa
s_{-}(z)s_{+}(w) &=& e^{-i\f{k}{N}H(z)} e^{i\f{k}{N}H(w)} \non
&\sim& (z-w)^{-(\f{k}{N})^2}[1- (z-w)\f{2k}{N} \partial H(z)]\\
{\tilde s}_{-}(z){\tilde s}_{+}(w) &=& e^{-i\f{k}{N}H({\bar z})}
e^{i\f{k}{N}H({\bar w})} \non
&\sim& ({\bar z} -{\bar w})^{-(\f{k}{N})^2} [1-({\bar z}-{\bar 
w})\f{2k}{N}{\bar \partial} H({\bar z})]
\eeqa

\noindent
For the bosonic twist fields,

\beqa
\s_{-}(z)\s_{+}(w)\sim |z-w|^{-2\f{k}{N}(1-\f{k}{N})}[1+\cdot\cdot\cdot]
\eeqa

Using these, the OPE of the compact part of $T_F$ with the twist fields
is,

\beqa
(\partial X {\bar \psi} + \partial {\bar X}
\psi)s_{+}\s_{+} \sim (z-w)^{-1}\tau_{+}e^{-i(1-\f{k}{N})H(z)} + 
\tau_{+}^{'}e^{i(1+\f{k}{N})H(z)}
\eeqa

This OPE includes higher twist operators and hence does not contain 
massless states which we are looking for.
The massless state is obtained from the ($k.\Psi k.{\tilde \Psi}$) term in 
the expansion (\ref{ope2}) with the other twist fields contracted.
The coupling for the term is,

\beqal{coupling1}
V_{\mu\nu}(k)=k_{\mu}k_{\nu}
\eeqa

This term is completely symmetric in the indices. It thus 
corresponds to the massless vertex for the graviton and the dilaton in 
the $(-1,-1)$ picture, which is the symmetric part of

\beqa
e^{-\phi}e^{-{\bar \phi}}\Psi_{\mu}{\tilde \Psi}_{\nu} e^{i(k+p).x}
\eeqa

The four point tachyon scattering amplitude with a massless graviton and 
dilaton exchange is given by,

\beqal{ope4pt1}
A_4 &=& 
V_{\mu\nu}(p_1)\f{1}{q^2}[\delta_{\mu\alpha}\delta_{\nu\beta}]V_{\alpha\beta}(p_3)\non
&=& 
p_{1\mu}p_{1\nu}\f{1}{q^2}[\delta_{\mu\alpha}\delta_{\nu\beta}]
p_{3\alpha}p_{3\beta}\non
&=& -\f{(p_1.p_3)^2}{s}=- \f{1}{4}\f{(u-2m^2)^2}{s}
\eeqa

where, $ q^2=-(p_1+p_2)^2=s $ and  $p_{i}^2=-m^2$. We 
have chosen the configuration of the 
momenta for the tachyon fields such that it matches with the original 
configuration used in (\ref{expect}) for convenience. Namely $p_1$ and 
$p_3$ 
corresponds to the momenta of the external $\phi^*$ field corresponding to 
the $V^-$ vertices. We still have to add the $t$-channel contribution to 
(\ref{ope4pt1}). Adding this we have,

\beqal{ope4pt}
A_4 &=&-\f{1}{4}[\f{(u-2m^2)^2}{s}+\f{(u-2m^2)^2}{t}]
\eeqa

This reproduces the poles which we have found in (\ref{string4pt}) as 
expected apart from a factor of $2\pi$ which comes in (\ref{string4pt}) 
from the integral over the vertex position.

\section{Amplitude from effective field theory}

In the previous section we have seen that the massless exchanges in the 
four point amplitude of the twisted sector tachyons are the graviton and 
the dilaton. In this section we calculate the tachyon four point amplitude  
with these massless exchanges, namely the graviton and dilaton from the 
effective 
field theory.

The action for the complex tachyon coupled to graviton and dilaton is 
given by,

\beqa
S=\f{1}{\kappa^2}\int d^Dx \sqrt{(-g)}[-R +
\f{4}{D-2}\Phi\partial^2\Phi +\f{1}{2}\phi^{*}(-\partial^2)\phi +
\f{1}{2}m^2e^{\f{4}{D-2}\Phi}\phi^{*}\phi]
\eeqa

Expanding $g_{\mu\nu}$ about the flat metric,

\beqa
g_{\mu\nu}=\delta_{\mu\nu}+\kappa h_{\mu\nu}
\eeqa

and rescaling the dilaton field by,

\beqa
\Phi \rightarrow \sqrt{\f{8}{D-2}}\Phi
\eeqa

We have,

\beqa 
S=\f{1}{\kappa^2}\int 
d^Dx[\f{1}{2}h_{\mu\nu}(-\partial^2+\cdot\cdot\cdot)h_{\mu\nu} 
- \f{1}{2}\Phi\partial^2\Phi +\f{1}{2}\phi^{*}(-\partial^2+m^2)\phi - 
h_{\mu\nu}T_{\mu\nu} + T\Phi]
\eeqa

The couplings of the graviton and dilaton to the complex scalar 
field are now,

\beqa
-\kappa h_{\mu\nu}T_{\mu\nu} \mbox{\hspace{.5cm} and \hspace{.5cm}  } 
\kappa T\Phi
\eeqa

Where,

\beqal{emt}
T_{\mu\nu}&=&-\f{1}{2}[\partial_{\mu}\phi^{*}\partial_{\nu}\phi+
\partial_{\nu}\phi^{*}\partial_{\mu}\phi] +\f{1}{2}\delta_{\mu\nu}
[\partial_{\alpha}\phi^{*}\partial_{\alpha}\phi + m^2|\phi|^2]\non
T&=&\sqrt{\f{2}{D-2}}m^2\phi^{*}\phi
\eeqa

The tachyon-graviton vertex and the graviton propagator in the harmonic 
gauge are given by,

\beqa
V_{\mu\nu}(p,k)=i\kappa[-\f{1}{2}(p_{\mu}k_{\nu}+p_{\nu}k_{\mu}) + 
\f{1}{2}\delta_{\mu\nu}(k.p -m^2)]
\eeqa

\beqa
\Delta_{\mu\nu\alpha\beta}(q^2)=\f{1}{q^2}[\delta_{\mu\alpha}\delta_{\nu\beta}+
\delta_{\mu\beta}\delta_{\nu\alpha} 
-\f{2}{D-2}\delta_{\mu\nu}\delta_{\alpha\beta}]
\eeqa

The four point amplitude for four massless scalar scattering with a 
graviton exchange is,

\beqa
A_4^{g} 
&=& 
V_{\mu\nu}(p_1,p_2)\Delta_{\mu\nu\alpha\beta}(q^2)V_{\alpha\beta}(p_3,p_4)\\
&=& 
\f{\kappa^2}{q^2}[(p_1.p_3)(p_2.p_4)+(p_1.p_4)(p_2.p_3)-(p_1.p_2)(p_3.p_4)\non 
&+& m^2(p_1.p_2+p_3.p_4)-\f{D}{D-2}m^4]\nn
\eeqa

Similarly, for the dilaton exchange we have,

\beqa
A_4^{d}=\f{\kappa^2}{q^2}\f{2}{D-2}m^4
\eeqa

Therefore, the four point tachyon amplitude with graviton and dilaton 
exchange is,

\beqa
A_4&=&A_4^{g}+A_4^{d}\non
&=&\f{\kappa^2}{q^2}[(p_1.p_3)(p_2.p_4)+(p_1.p_4)(p_2.p_3)-(p_1.p_2)(p_3.p_4)
+m^2(p_1.p_2+p_3.p_4)-m^4]\non
&=& 
-\f{\kappa^2}{4s}[(u-2m^2)^2+(t-2m^2)^2 - s^2]\\
&=&-\f{\kappa^2}{2}[\f{(t-2m^2)^2}{s}+t - 2m^2]
\eeqa

\noindent
In writing (58) from the previous step, we have used $p_i^2=-m^2$ 
and put in $q^2=(p_1+p_2)^2=-s$. From (58) to (59) we have used
$s+t+u=4m^2$ which uses the mass shell conditions. 
Now including the $t$-channel process we get,

\beqal{field4pt}
A_4=-\f{\kappa^2}{2}[\f{(t-2m^2)^2}{s}+\f{(s-2m^2)^2}{t} 
+ (t+s) - 4m^2]
\eeqa

Comparing with the pole term of the string amplitude (\ref{string4pt}), 
we see that the pole is due to graviton exchange. 
This also relates
$\kappa$ to $g_c$ which is found to be,

\beqa
\f{\kappa^2}{2}=4\pi^2g_c^2
\eeqa

\noindent
After subtraction, the nonderivative quartic term and the derivative 
terms left behind are, 

\beqa
-4\pi^2g_c^2[-4m^2+2(s+t)]
\eeqa

\subsection{The Potential}

We can now write down the potential for the tachyon upto the quartic term. 
It may be noted that the sign of the quartic term is to be fixed relative 
to the sign of the pole terms which has to be positive. So we get the 
quartic coupling as,

\beqa
\lambda &=& (4\pi^2g_c^2) \times (-4m^2)\non
\eeqa 

which is positive since $m^2=-k/N$. The tachyon potential is now,

\beqa
V(\phi^*\phi)&=&\f{1}{2}m^2(\phi^*\phi)+\f{\lambda}{4}(\phi^*\phi)^2 
\non
&=& -\f{k}{2N}(\phi^*\phi)+4\pi^2g_c^2\f{k}{N}(\phi^*\phi)^2
\eeqa

The potential has a minimum at $|\phi|^2=1/(16\pi^2g_c^2)$ which is of 
$O(1)$. We expect the 
nearest minimum to correspond to the $C/Z_{N-k}$ orbifold. In this case
the height of the potential from  $C/Z_{N}$ to $C/Z_{N-k}$ is given by,

\beqal{height}
\Delta=\f{1}{64\pi^2g_c^2}\f{k}{N}
\eeqa

One may compare this, upto normalizations, with the conjectured height 
\cite{atish2} which is,

\beqa
\Delta=4\pi(\f{1}{N-k}-\f{1}{N})\sim 4\pi \f{k}{N^2}
\eeqa

This shows that the perturbative result (\ref{height}) is off by a factor 
of $1/N$. At 
this point we are not in a position to trust this perturbative result. 
Higher point amplitudes will most likely modify this.

There are  various indications that higher point interaction terms 
in the amplitude are in fact important and are not of order less than 
$1/N$. We may note that with a potential upto the quartic coupling having 
global $O(2)$ symmetry,
in the spontaneously broken theory there is a massless scalar 
corresponding to the Goldstone boson and a massive particle of mass, 
$-2m^2=2k/N$. This means that in the spectrum of $C/Z_{N-k}$ to which the  
theory is supposed to flow, there must be a massless and a massive scalar  
of mass $2k/(N-k)(\sim 2k/N+\cdot\cdot\cdot;\mbox{for large $N$})$. 
However the 
spectrum does not contain these scalars.  
The absence of the massless Goldstone particle indicates that the O(2) 
symmetry of the tachyon potential has to be broken. This can only happen 
if the correlation functions of $N$ twist operators are also of order 
$1/N$.  

This fact may also seen as follows. The three point graviton vertex 
and the two tachyon and one graviton vertex, both have two powers of 
momentum (Figure 1). The four point tachyon amplitude with a graviton 
exchange has 
two positive powers of momentum.

\begin{figure}[h]
\begin{center}  
\epsfig{file=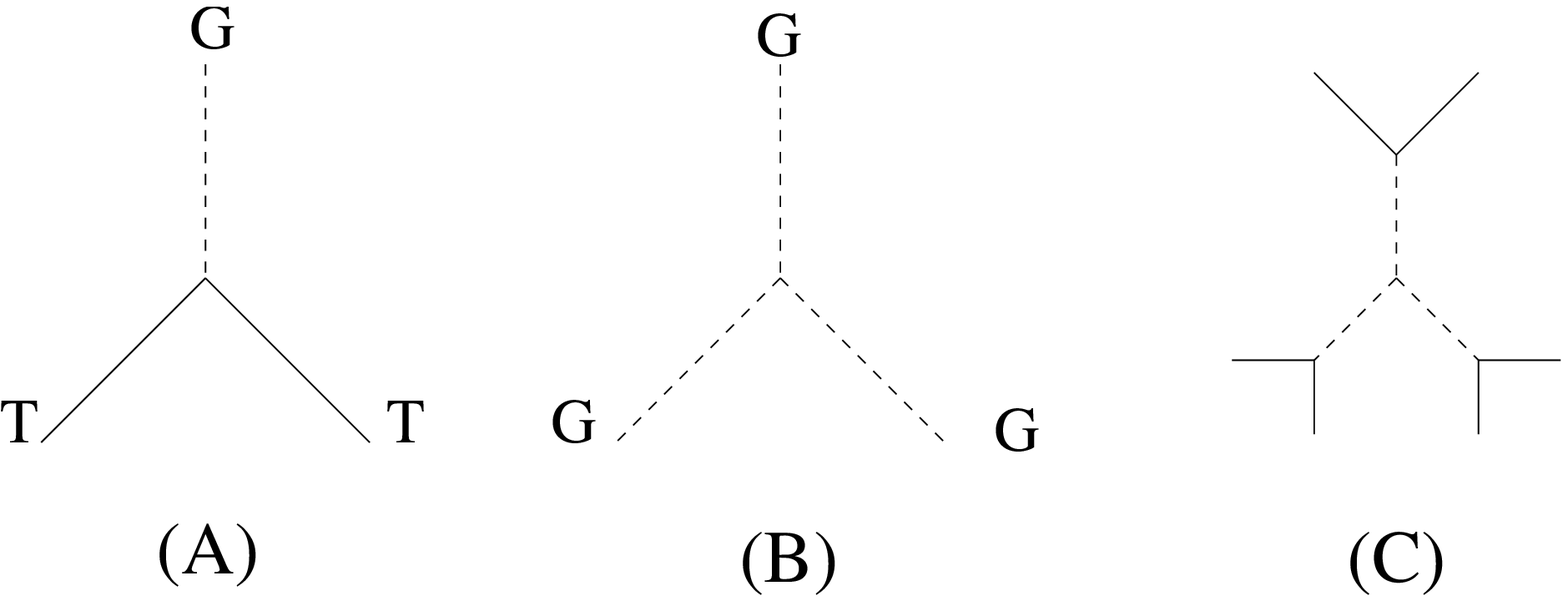, width= 8 cm,angle=0}
\vspace{ .1 in }
\begin{caption}
{(A)Two tachyon one graviton vertex, (B) Three graviton vertex, (C) 
Six-point tachyon amplitude with graviton exchange.}
\end{caption}
\end{center}
\end{figure}

With the addition of two more external tachyons, using (A), the positive 
power of momentum for the six point amplitude (C), remains two. Of course 
with a three graviton vertex insertion, we can introduce more negative 
powers, but there are tree diagrams where we can 
have two positive powers 
of momentum. The $N$ point contact term is obtained by subtracting these 
graviton exchanges from the full $N$ point tachyon amplitude from string 
theory. The graviton exchange diagrams as mentioned contains two 
positive powers of momentum, which in the onshell limit give terms 
proportional to $m^2\sim O(1/N)$.
It is thus very likely that the subtracted term  
would give a $1/N$ dependence as the leading part.

We further see that the dilaton field redefinitions 
such as, 

\beqa
\Phi \rightarrow \Phi + c \phi^*\phi
\eeqa

where $c$ is a constant, can change the value 
of the contact term and can even change the sign. The minimum thus depends 
crucially on the expectation value of the dilaton field which when becomes 
large, makes this perturbative analysis anyway unreliable. 
A similar observation was made in \cite{dine}, in the context of closed 
string tachyon condensation with Rohm's Compactification.
In general the 
existence of the minimum is independent of field redefinitions. The fact 
that we are able to change the nature of the potential by field 
redefinitions implies the potential upto the quartic term does not shed 
much light on the minimum of the theory.
It is argued in various approaches that the Type II 
theory on the $C/Z_N$ orbifold ultimately upon closed string tachyon 
condensation goes to the Type II theory on flat space. Our analysis does 
not give a proof of this observation. If we assume this to be true, that a 
stable minimum exists, then following the above arguements, we may 
conclude that the higher point terms are indeed necessary and are of the 
order $1/N$.  

\section{Conclusion}

We have studied the condensation of closed string tachyons for Type II 
strings on the $C/Z_N$ orbifold. We constructed the potential for the 
tachyons upto the quartic term in the large $N$ limit by subtracting the 
massless exchanges from the four point tachyon amplitude computed from 
string theory. We expect the minimum of the potential for  
the near marginal tachyons for the $k$-th twisted
sector, in the large $N$ limit to correspond to the 
$C/Z_{N-k}$ orbifold. When compared to the conjectured value for the 
height 
of the potential for the $C/Z_N$ orbifold, we find a mismatch by a factor 
of $1/N$. However, we have argued that the higher point amplitudes are 
indeed important and are of the same order in $1/N$ as the quartic term. 
A potential upto the quartic term after spontaneous symmetry breaking 
gives masses which are not there in the spectrum for closed string on  
$C/Z_{N-k}$ to which the $C/Z_N$ theory is expected to flow. This leads us 
to conclude that the higher point amplitudes including the global $O(2)$ 
breaking term, $\phi^N$, must all be of order $1/N$ so that the potential 
gives a mass spectrum, consistent with that of the $C/Z_{N-k}$ 
orbifold. We have also argued that field redefinitions can alter the 
contact term and can even change the sign. 
If the theory can be deformed so that the minimum can be reliably reached 
in perturbation theory, then, a direct approach such as the one discussed 
in this paper can ascertain whether this minimum has the required properties.
\\
\\
\noindent
{\bf Acknowledgments :} We would like to thank A. P. Balachandran for 
useful discussions. 
\\
\\

\noindent
{\bf Note added :} The computation of the quartic coupling for the twisted
sector tachyon is done recently by Atish Dabholkar, Ashik Iqubal and Joris
Raeymaekers in {\it Off-Shell Interactions for Closed-String Tachyons}
[hep-th/0403238]. It is pointed out correctly by these authors that
additional contributions to the four point contact term will also come
from the massive untwisted states with momentum along the orbifold plane,
$C/Z_N$. This modifies the $1/N$ dependence of the quartic term which we
computed in this paper to a more suppressed $1/N^3$ dependence. However
with this modification, the expectation that the large $N$ approximation
may may be used to study the RG flow due to tachyon condensation from the
$C/Z_N$ orbifold to lower nonsupersymmetric orbifolds is even further
weakened. The conclusions of this paper thus remain unaltered.

\end{document}